\documentclass[prb,prb,reprint,aps,showpacs,showkeys,twocolumn]{revtex4-1}

\pdfoutput=1

\usepackage[utf8]{inputenc}
\usepackage[T1]{fontenc}
\usepackage{hyperref}
\usepackage{color}
\usepackage{graphicx}
\usepackage{units}
\usepackage{amsmath}

\hypersetup{colorlinks = true, citecolor = blue, breaklinks = true}

\begin{document}

\title{Effect of epitaxial strain on cation and anion vacancy formation in MnO}
\date{\today}
\author{Ulrich Aschauer}
\affiliation{Materials Theory, ETH Zurich, Wolfgang-Pauli-Strasse 27, CH-8093 Z\"urich, Switzerland}
\author{Nathalie Vonr\"uti}
\affiliation{Materials Theory, ETH Zurich, Wolfgang-Pauli-Strasse 27, CH-8093 Z\"urich, Switzerland}
\author{Nicola A. Spaldin}
\affiliation{Materials Theory, ETH Zurich, Wolfgang-Pauli-Strasse 27, CH-8093 Z\"urich, Switzerland}

\begin{abstract}
Biaxial strain in coherent epitaxial thin films can have a pronounced effect on the point defect profile in the film material. Detailed fundamental knowledge of the interaction of strain with point defects is crucial in understanding the stoichiometry and resulting properties of strained thin films. Here we investigate the effect of biaxial strain on the formation energy of cation and anion vacancies using MnO as a model system. Our density functional theory calculations show that, as expected from local volume arguments, compressive strain favours the formation of cation vacancies. Interestingly, we find that small compressive and tensile strains lead to ordering of the resulting holes along the in-plane and normal direction respectively, which should manifest in different anisotropic properties in the two strain states.
\end{abstract}

\maketitle

\section{Introduction}

Point defects in ionic materials are ubiquitous and often strongly affect or even determine the material's properties. They can be classified as extrinsic, such as impurities or dopants, or as intrinsic, when the crystal lattice is locally imperfect for example by missing, displaced or interchanged atoms. Intrinsic defects occur even in the purest of materials as their formation is thermally activated. The relative concentration of intrinsic point defects is thus a function of the Arrhenius exponential factor of their formation energy $\Delta G_\mathrm{form}$, the exact form depending on the respective defect equilibrium \cite{Kroger:1956bl}. As the number of atoms and electrons may change during defect formation, the formation energy depends on the chemical potentials of the atoms and electrons in their respective reservoirs \cite{Freysoldt:2014ej} in addition to the internal energy cost to form the defect. This leads to the well-established effect of the environment on the defect profile of a material via the partial pressure of the constituent elements (see e.g. Ref \onlinecite{Ertekin:2012ez}).

We have recently shown using first principles density functional calculations that besides a reduced oxygen partial pressure in the surrounding environment, oxygen vacancy formation in manganite perovskites is also facilitated by tensile strain, which can, for example, be applied using coherent heteroepitaxy in a thin film \cite{Aschauer:2013hf}. Enhanced oxygen deficiency under tensile strain was indeed experimentally observed in films of La$_{0.7}$Ca$_{0.3}$MnO$_3$ \cite{Kawashima:2015hy} as well as in cobaltite perovskites \cite{Biskup:2014eo}. The effect of strain on the formation of oxygen vacancies can be seen as the converse of the concept of chemical expansion \cite{Adler:2001ww}: Depending on the charge of a point defect relative to the ideal lattice site, its formation is accompanied by oxidation or reduction of neighbouring transition metal ions, which lead to changes in ionic radius and manifest macroscopically as changes in lattice parameters. Conversely, if one imposes lattice parameters, for example by epitaxial strain, the locally expanded/contracted volume will facilitate the reduction/oxidation of transition metal ions, leading to a larger concentration of point defects, corresponding to this changed oxidation state. We have further shown that epitaxial strain has the effect of ordering point defects on inequivalent sites in the lattice, with tensile strain strongly favouring oxygen vacancy sites for which the broken bond axis lies within the strained plane.

In this work we study the response of the cation (manganese) vacancy profile to epitaxial strain, which is expected to be favoured under compression. For completeness we also study the strain response of anion (oxygen) vacancies, which from volume arguments should be favoured under tensile strain. We choose the rock salt oxide MnO as our model system because its $\mathrm{Mn^{2+}}$ cations are readily oxidised and it is structurally simpler than related perovskite oxides. Our main findings are that, as expected, cation vacancies are favoured by compressive strain, with strain additionally inducing an ordering of the holes on sites around the defect. Against our expectation anion vacancies are not favoured by tensile strain but rather also by compressive strain, which stems from the formation of an F-center instead of charge localisation on the transition metal site.

\begin{figure}
\includegraphics[width=\columnwidth]{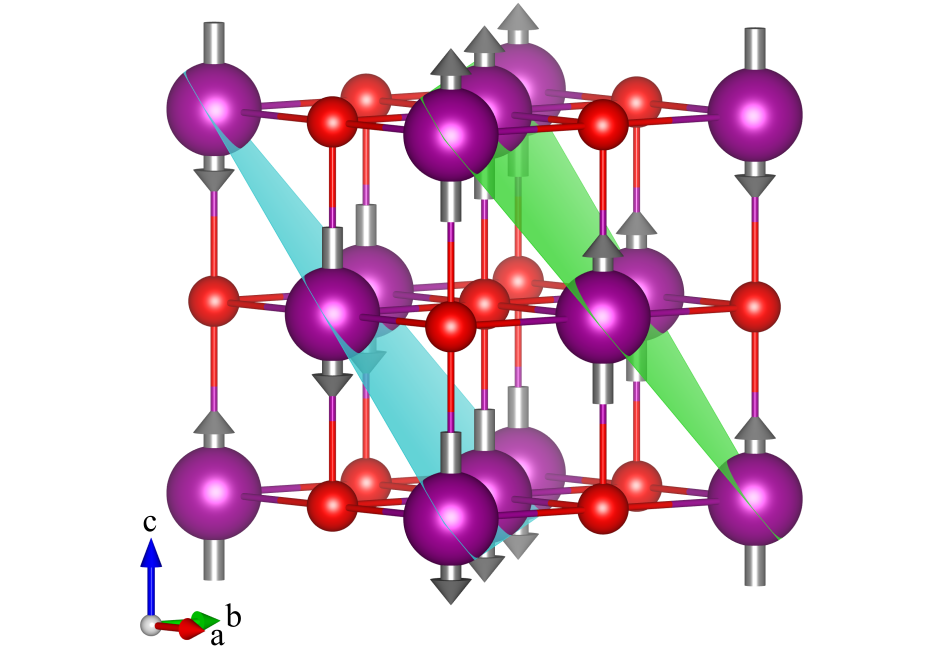}
\caption{\label{fig:structure}The cubic rocksalt unit cell of MnO. Arrows show the direction of magnetic moments on the Mn atoms. In the AFM-II phase moments of the same spin form (111) layers, two of which are shown with the green and blue planes. Color code: Mn=purple (large), O=red (small).}
\end{figure}

Above the N\'{e}el temperature of 118 K MnO is a paramagnetic insulator with the ideal cubic rock-salt crystal structure. Below this temperature the transition metal magnetic moments order in the antiferromagnetic (AFM) type II arrangement shown in Fig. \ref{fig:structure}, which consists of ferromagnetic $\left(111\right)$ planes, which align antiferromagnetically with each other \cite{Roth:1958gc}. The exchange interactions in this magnetic phase induce a magnetic stress \cite{Filippetti:2000wa} leading to a compression along the $\left[111\right]$ direction \cite{Pask:2001kq,Schron:2012fs}, which manifests itself as a rhombohedral distortion of the unit cell. Experimentally reported distortion angles range from 0.43$^\circ$ to 0.62$^\circ$ \cite{Roth:1958gc,Cheetham:1983kf,Shaked:1988do}.

The main point defects occurring in MnO are known to be Mn vacancies ($\mathrm{V_{Mn}}$)\cite{Hed:1967hf,Kofstad:1983bu}. Using electrical conductance measurements, the charge state of these $\mathrm{V_{Mn}}$ was determined to be the neutral vacancy ($\mathrm{V_{Mn}^{''}}$ in Kr\"oger-Vink notation) at low oxygen partial pressure ($\mathrm{p_{O_2}}$) and low defect concentrations \cite{Hed:1967hf}. At intermediate $\mathrm{p_{O_2}}$ the charge state changes to $\mathrm{V_{Mn}^{'}}$ and finally to $\mathrm{V_{Mn}^{x}}$ at high $\mathrm{p_{O_2}}$. The predominance of $\mathrm{V_{Mn}^{''}}$ at low $\mathrm{p_{O_2}}$ was later confirmed using ionic transport measurements, but a defect cluster of four $\mathrm{V_{Mn}^{''}}$ and one Mn interstitial ($\mathrm{Mn_i^{\bullet\bullet}}$) better fits the ionic transport data at high $\mathrm{p_{O_2}}$ \cite{Kofstad:1983bu}. A theoretical study of $\mathrm{V_{Mn}}$ in MnO suggested that the O atoms around the vacancy acquire a magnetic moment and that the moments of the next nearest neighbour Mn ions are reduced compared to bulk-like Mn ions \cite{Kodderitzsch:2003bi}. These calculations were carried out in a 64-atom cubic supercell however, which, as will be discussed below, may lead to spurious results. Recently it was predicted that $\mathrm{V_{Mn}}$ at concentrations higher than 20\% may render MnO a room temperature ferrimagnetic material \cite{Seike:2014ks}.

\section{Computational Methods}

Our density functional theory (DFT) calculations were performed with the PBEsol exchange-correlation functional \cite{Perdew:2008fa} using the VASP code \cite{Kresse:1993ty,Kresse:1994us,Kresse:1996vk,Kresse:1996vf}. Wave functions were expanded in plane waves up to a kinetic energy cutoff of 550 eV. We used projector augmented wave (PAW) potentials \cite{Blochl:1994uk,Kresse:1999wc} with Mn(3p, 3d, 4s) and O(2s, 2p) electrons in the valance. The rhombohedral distortion in MnO is severely overestimated by local (LDA) and semi-local (GGA) density functionals, the DFT+U method \cite{Anisimov:1991wt} applied to the Mn d orbitals resulting in a much improved agreement with experiment \cite{Schron:2012fs}. Within this method, structural changes occur in the range U=0-2 eV whereas larger values of U do not further change the structure \cite{Schron:2012fs}. In our calculations we applied U=4.0 eV to the Mn 3d orbitals, which also results in correct redox energetics in MnO \cite{Wang:2006kn}. Reciprocal space for the rhombohedral 4 atom (a,b,c=5.362 \AA, $\alpha$, $\beta$,$\gamma$=34.258$^\circ$) antiferromagnetic (AFM-II) unit-cell was sampled using a $\Gamma$-centred 7x7x7 grid. Oxygen vacancy formation was studied in a 64 atom (a,b,c=8.865 \AA, $\alpha$,$\beta$,$\gamma$=90.857$^\circ$) supercell where k-space was sampled using a 4x4x4 $\Gamma$-centred mesh. Cation vacancies were studied in a 512 atom (a,b,c=17.730 \AA, $\alpha$,$\beta$,$\gamma$=90.857$^\circ$) supercell with only the $\Gamma$ point used for reciprocal-space sampling.

For strained film calculations we imposed the epitaxial constraint by setting one of the three cell angles to $90^\circ$, constraining the two in-plane lattice parameters and relaxing the length and direction of the lattice vector normal to the substrate at each value of imposed epitaxial strain. Internal coordinates in all calculations were relaxed until forces converged below 0.001 eV/\AA.

For cation vacancies multiple competing localised hole arrangement patterns exist. We achieved a systematic study of their relative energies by imposing different patterns via the occupation matrix of the $\mathrm{Mn^{3+}}$ atoms surrounding the vacancy \cite{Allen:2014ft}. The lattice was first relaxed with these constraints to obtain the lattice distortion corresponding to the desired hole order. The occupation constraints were then removed but the hole localisation pattern was retained due to the  lattice distortion.

Defect formation energies were computed according to equation \ref{eq:formation} \cite{Zhang:1991ck,Freysoldt:2014ej}:
\begin{equation}
\Delta E_f = E_\mathrm{def} - E_\mathrm{stoi} - \sum_{i}{n_i\mu_i}+qE_\mathrm{fermi}+E_\mathrm{corr}.
\label{eq:formation}
\end{equation}
Here $E_\mathrm{def}$ and $E_\mathrm{stoi}$ are the DFT total energies of the defective and stoichiometric supercells respectively at a given level of epitaxial strain. For charged defects ($q\neq0$), a neutralising background charge was applied and the formation energy corrected by the Fermi energy ($E_\mathrm{fermi}$, the chemical potential of the electrons) and $E_\mathrm{corr}$, a correction derived using the sxdefectalign code \cite{Freysoldt:2009ih,Freysoldt:2010gx} to align the electrostatic potentials of the stoichiometric and charged defective cell. The Fermi energy is defined relative to the valence-band edge and can take values from zero up to the band-gap of the material.

In this work we denote defects and their charge states using Kr\"oger-Vink notation \cite{Kroger:1956bl}, which is of the following form: $\mathrm{M_S^C}$. In this notation M refers to the defect species, which for the vacancies we consider here is V. The subscript S indicates the lattice site the defect occupies and can be either a species of the perfect lattice or an interstitial denoted as i. The superscript C refers to the charge state of the defect \textit{relative} to the lattice site it occupies. A charge of zero is indicated by $\mathrm{x}$ and positive and negative elementary charges by $\mathrm{\bullet}$ and $\mathrm{'}$ respectively. Creating a Mn vacancy without adjusting the charge (q=0 in Eq. \ref{eq:formation}) leads to a neutral vacancy on a $\mathrm{Mn^{2+}}$ site, which thus has a \textit{relative} charge of -2e, which in Kr\"oger-Vink notation is $\mathrm{V_{Mn}^{''}}$. This can also be regarded as the Mn vacancy introducing two holes into the material. Adding electrons to the system (q=1 and q=2 in Eq. \ref{eq:formation}) will change the relative charge state of the vacancy to $\mathrm{V_{Mn}^{'}}$ and $\mathrm{V_{Mn}^{x}}$ respectively, which means that there is one hole or no hole respectively introduced into the material.

During vacancy formation, $n_i$ atoms of a given species $i$ are removed from the system, which is taken into account by their chemical potential $\mu_i$. The chemical potentials of Mn and O can vary within a range limited by the decomposition of MnO to metallic Mn ($\mathrm{\mu_{Mn}=\mu_{Mn,metal}}$) on one side and the condensation of $\mathrm{O_2}$ ($\mathrm{\mu_O=\frac{1}{2}\mu_{O_2}}$) on the other side. The chemical potentials are related by $\mathrm{\mu_{Mn}+\mu_O=\mu_{MnO,bulk}}$, and we express $\mathrm{\mu_{Mn}}$ as a function of $\mathrm{\mu_{O}}$ in this work.

\section{Results \& Discussion}

\subsection{Mn vacancies in bulk MnO}

\begin{figure}
\includegraphics[width=\columnwidth]{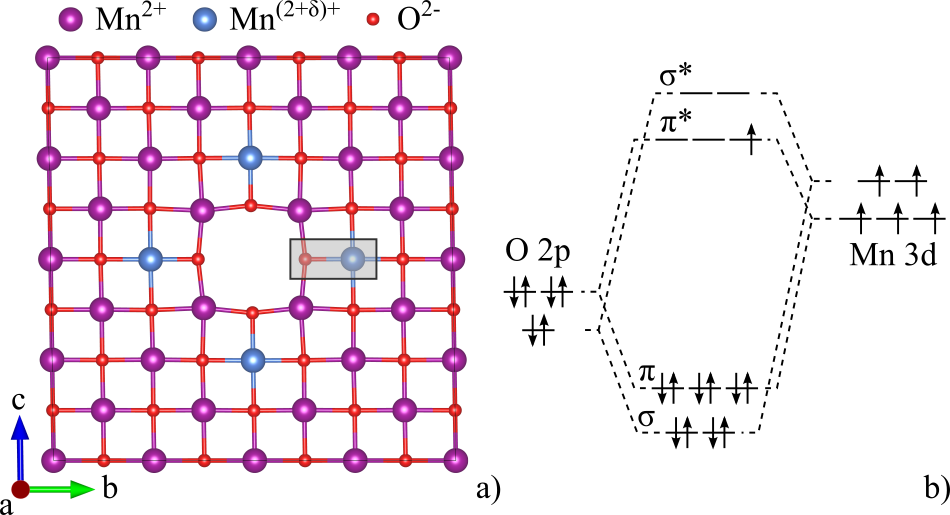}
\caption{\label{fig:VMn_schematic}a) Computed atomic structure showing the relaxation of the O atoms around a $\mathrm{V_{Mn}^{x}}$ vacancy. The relaxation leads to a significant shortening of the Mn-O bonds towards the Mn ions highlighted in blue. b) Schematic molecular orbital diagram for the Mn-O bond highlighted in panel a). The bond shortening  increases the energy of the $\pi^*$ molecular orbital, which becomes a preferential site for hole accommodation.}
\end{figure}

To analyse the electronic structure of a cation vacancy in MnO, we start with the simplest case: the $\mathrm{V_{Mn}^x}$ vacancy (which does not introduce holes in the lattice) in unstrained bulk MnO. In Fig. \ref{fig:VMn_schematic}a) we show our calculated two-dimensional section through such a $\mathrm{V_{Mn}^x}$. The oxygen atoms surrounding the vacancy relax away from the vacancy, leading to shortened Mn-O bonds between the relaxing O ions and six Mn ions, four of which are visible and highlighted in blue in Fig. \ref{fig:VMn_schematic}a). This leads to an enhanced overlap and hence a higher energy of the $\pi^*$ molecular orbitals formed between the $\mathrm{t_{2g}}$ Mn 3d and the O 2p orbitals as schematically shown in Figure \ref{fig:VMn_schematic}b). These states above the valence-band edge in the DOS have primarily Mn $t_{2g}$ character and are localised on the highlighted Mn ions. Due to their higher energy these states accommodate holes more readily than the corresponding molecular orbitals on regular Mn ions without shortened bonds.

We find that larger Hubbard U values applied to the Mn d orbitals lower the energy of these occupied anti-bonding states, making their character gradually less Mn-like.  We expect this to lead to a transition from well-localised polaronic holes to more delocalised band-like behaviour. We have determined however that unphysically large values of U (> 8 eV) are required to induce a significant change in band-edge character, which implies that the picture of well localised holes is valid in Mn deficient MnO.

This ordering of holes around the vacancy implies that in order to fit even only one bulk-like Mn between Mn vacancies one needs a supercell with at least four cubic rock-salt MnO units along each of the crystal axes. This requires us to use a 512 atom 2x2x2 supercell of the cubic 64 atom AFM-II unit cell. Previous work using smaller 64-atom cells \cite{Kodderitzsch:2003bi} was likely affected by the fact that these hole-accomodating Mn ions along one direction were equivalent by periodic boundary conditions. This prevents charge localisation as the hole count is artificially doubled on a given Mn site. Indeed, if we perform computations in the smaller 64-atom cell, we recover the delocalised half-metallic state reported reported in Ref. \onlinecite{Kodderitzsch:2003bi}.

\begin{figure}
\includegraphics[width=\columnwidth]{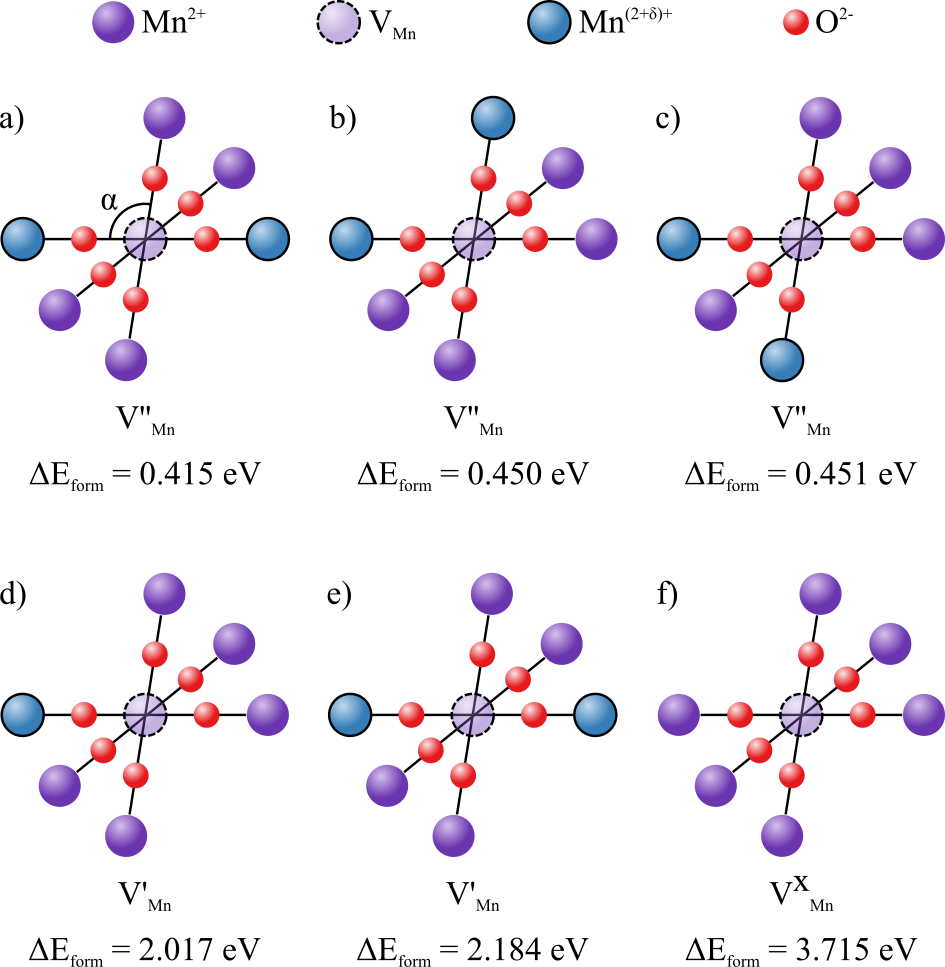}
\caption{\label{fig:VMn_bulk} Mn vacancy formation energies (ambient conditions: $\mu_O$ = -4.85 eV, $\mathrm{E_{fermi}}$ = 2.8 eV) in bulk MnO for different charge states and arrangements of holes. Mn atoms accommodating holes are highlighted in blue. Note that the rhombohedral distortion as indicated by the angle $\alpha$ in panel a) is exaggerated for clarity. $\mathrm{V_{Mn}^{''}}$ with two holes on a) two opposite Mn, b) two Mn in an angular arrangement with an angle $>\frac{\pi}{2}$ and c) with angle $<\frac{\pi}{2}$. For $\mathrm{V_{Mn}^{'}}$ the single hole can be localised on d) one Mn or e) partially distributed over two opposite Mn while in f) for $\mathrm{V_{Mn}^{x}}$ there is no hole.}
\end{figure}

Depending on the charge state of the cation vacancy, between zero and two holes are created. The state with two holes ($\mathrm{V_{Mn}^{''}}$) leads to the lowest occupation of empty high-energy orbitals and is thus expected to be energetically most favourable compared to the $\mathrm{V_{Mn}^{'}}$ and $\mathrm{V_{Mn}^{x}}$ with only one or no hole respectively. This is indeed what we see in Fig. \ref{fig:VMn_bulk} where we report the lowest formation energy for a $\mathrm{V_{Mn}^{''}}$ to be 0.415 eV (Fig. \ref{fig:VMn_bulk} a), whereas the formation energies for $\mathrm{V_{Mn}^{'}}$ (Fig. \ref{fig:VMn_bulk} d) and $\mathrm{V_{Mn}^{x}}$ (Fig. \ref{fig:VMn_bulk} f) are much higher at 2.017 eV and 3.715 eV respectively. These formation energies are computed for a Fermi energy of 2.8 eV but the situation does not change throughout the whole accessible range of the Fermi energy.

Since there are a maximum of two holes but six possible Mn sites to accommodate them, different patterns of distributing the holes are possible. We also considered states where one hole was distributed over multiple of the six possible Mn sites, which we refer to as \textit{distributed} arrangements. We attempted to stabilise all likely combinations and were able to achieve the patterns shown in Fig. \ref{fig:VMn_bulk}; other cases decayed into one of these patterns. In all cases the holes remained localised on the Mn sites highlighted in Fig. \ref{fig:VMn_schematic}a) forming polaronic states with corresponding lattice distortion. While the undercoordinated O atoms relax away from the vacancy, shortening the Mn-O bonds from 2.21 \AA\ in the bulk to 2.11 \AA\ as shown in Fig. \ref{fig:VMn_schematic}a), the bonds which accommodate holes are additionally shortened to 1.92 \AA.

For the $\mathrm{V_{Mn}^{''}}$ defect we see slight variations in the formation energy depending on the hole pattern. The most favourable hole arrangement pattern is when the two holes are on opposite sides of the vacancy (Fig. \ref{fig:VMn_bulk} a). Due to the slight rhombohedral distortion, two different angular arrangements of holes exist - one where the two hole-carrying Mn lie at an angle of more than 90 degrees with respect to the vacancy (Fig. \ref{fig:VMn_bulk} b) and one where this angle is smaller than 90 degrees (Fig. \ref{fig:VMn_bulk} c). We find that the situation with the larger angles is favoured by a minute amount, consistent with the electrostatic interaction between like-charge holes. Regarding the less stable $\mathrm{V_{Mn}^{'}}$ defects, we note that hole distribution (half a hole on two opposite Mn) shown in Fig. \ref{fig:VMn_bulk} e) is less stable than the localised state (one hole on a single Mn) shown in Fig. \ref{fig:VMn_bulk} d). Hole distribution thus seems unfavourable, which is in agreement with the results for $\mathrm{V_{Mn}^{''}}$, where all distributed patterns were found to be unstable and decay to fully localised orderings.

From the above results on the formation energy of differently charged $\mathrm{V_{Mn}}$ in bulk MnO it appears that the $\mathrm{V_{Mn}^{''}}$ has by far the lowest formation energy at all possible values of the Fermi energy. This is in agreement with experimental findings at low $\mathrm{p_{O_2}}$; the defect clusters reported to form at higher $\mathrm{p_{O_2}}$ \cite{Kofstad:1983bu} were not considered in the present work.

\subsection{Mn vacancies in strained MnO films}

\begin{figure}
\includegraphics[width=\columnwidth]{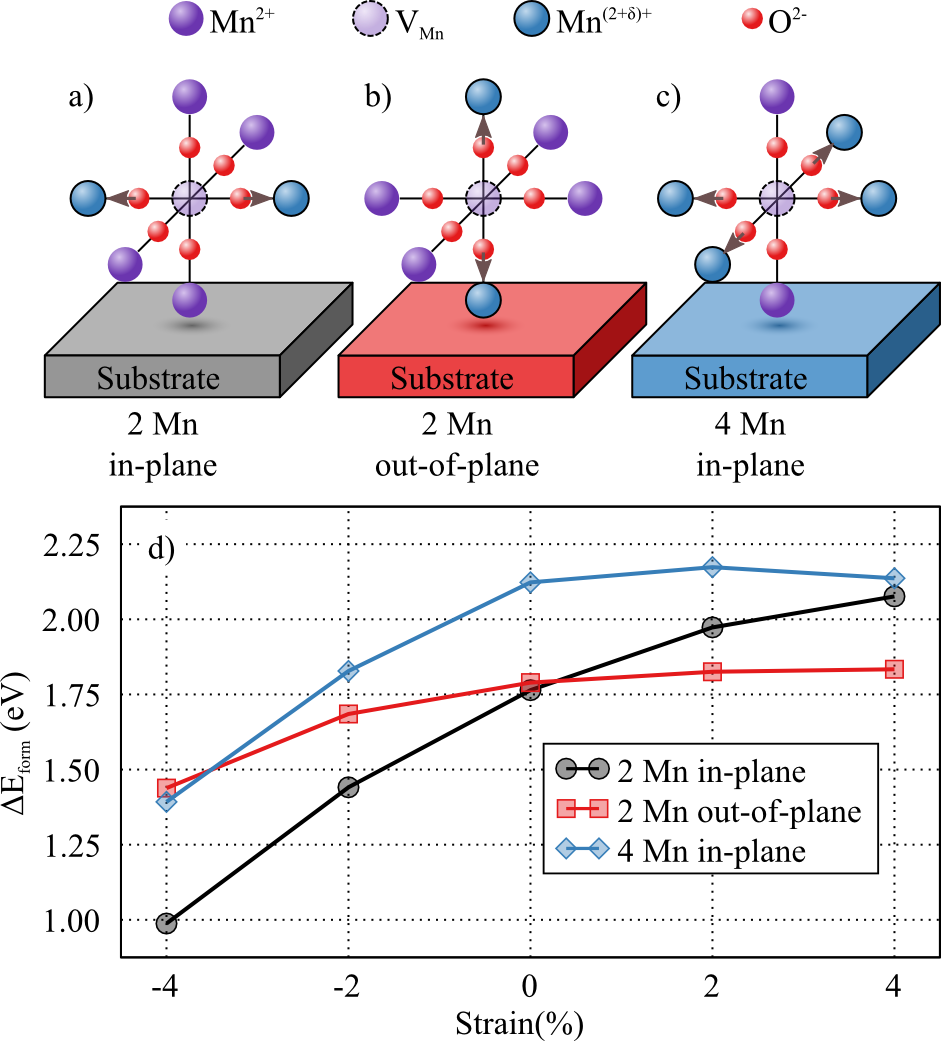}
\caption{\label{fig:VMn}Different likely arrangements of holes for $\mathrm{V_{Mn}^{''}}$ (q=0) vacancy formation in a strained thin film: a) one hole on each of two opposite in-plane Mn, b) holes on two opposite out-of-plane Mn and c) partially distributed holes on four in-plane Mn. In panel d) the formation energies for these hole arrangements are shown.}
\end{figure}

Having established the electronic structure of the Mn vacancy in bulk MnO, we will now consider strained thin films but restrict our analysis to $\mathrm{V_{Mn}^{''}}$ defects, which, as established above, have significantly lower formation energies than the other charge states. In a strained thin film, the likely hole arrangement patterns become more complex than in the bulk material, as strain lifts the symmetry between the in-plane and out-of-plane Mn sites around the defect. Out of the resulting possibilities for hole ordering, we find that the partially distributed state with holes on two in-plane and two out-of-plane Mn sites is unstable as is the distributed arrangement over six Mn sites. The remaining stable configurations are schematically shown in Figures \ref{fig:VMn}a) to c). From their respective formation energies shown in Figure \ref{fig:VMn}d) we see that the partially distributed hole arrangement on four Mn ions is less favourable than the two localised hole arrangements. The lowest energy hole arrangement is thus one hold on each of two Mn through the whole range of strains but switches from an in-plane pattern under compressive strain to an out-of-plane pattern under tensile strain. This transition could be measurable as a different anisotropy in the properties of Mn deficient MnO films under slight tensile and compressive strain.

\subsection{Oxygen vacancies}

\begin{figure}
\includegraphics[width=\columnwidth]{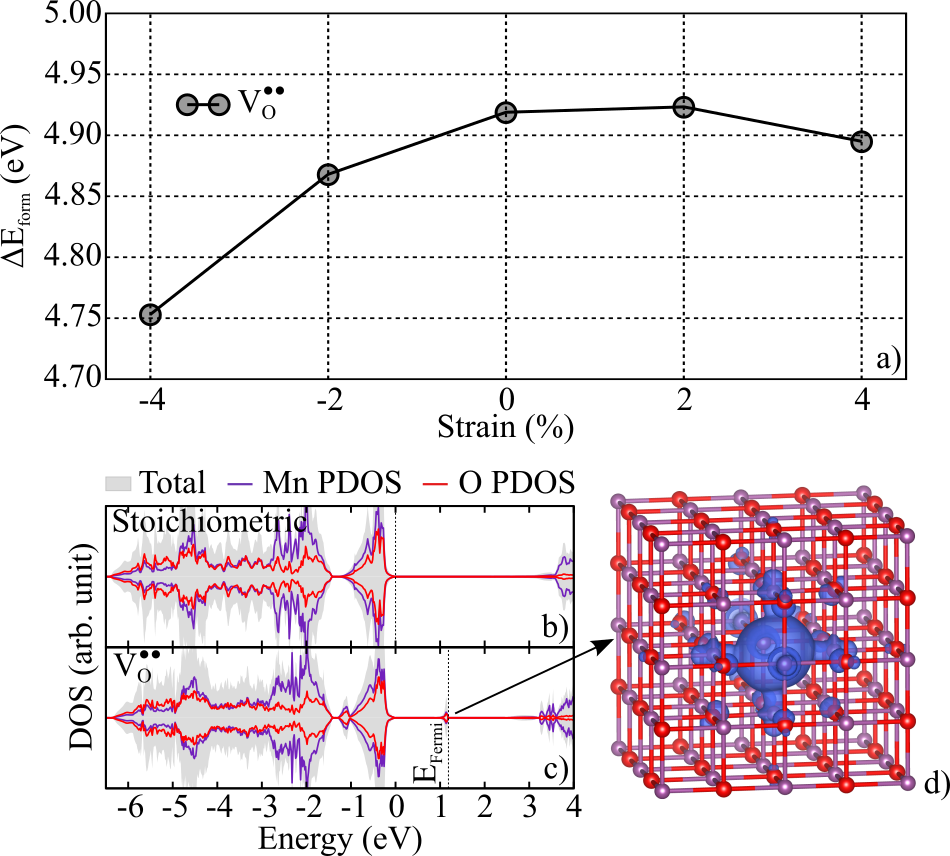}
\caption{\label{fig:VO}a) Oxygen vacancy formation energy (ambient conditions: $\mu_O$ = -4.85 eV) as a function of strain for the charge neutral ($\mathrm{V_O^{\bullet\bullet}}$, q=0) oxygen vacancy. Electronic densities of states for b) stoichiometric MnO, c) a $\mathrm{V_O^{\bullet\bullet}}$ oxygen vacancy with two electrons forming a defect state 1.2 eV above the valence band edge. The DOS have been aligned at the valence band edge and the Fermi energy is indicated by the dashed line. The 0.004 e/\AA$^3$ isosurface in panel d) shows the charge density associated with the F-center defect state for the $\mathrm{V_O^{\bullet\bullet}}$.}
\end{figure}

Despite the fact that oxygen vacancies have limited experimental relevance in MnO, we have studied the strain response of their formation energy for completeness and to see if it follows the local volume variation model \cite{Aschauer:2013hf}. Looking at the formation energy of the $\mathrm{V_O^{\bullet\bullet}}$ defects shown in Figure \ref{fig:VO}a) we see that 4\% tensile strain very slightly favours oxygen vacancy formation. There is however a much larger lowering of the formation energy observed for compressive strain, which is in sharp contrast to the behaviour expected for defect electrons localised on transition metal sites \cite{Aschauer:2013hf}.

From the electronic density of states we see by comparing the stoichiometric case (Fig. \ref{fig:VO}b) with that for $\mathrm{V_O^{\bullet\bullet}}$ (Fig. \ref{fig:VO}c), that oxygen vacancy formation leads to the appearance of a two-electron defect state about 1.2 eV above the valence band edge. The isosurface of the charge density associated with this defect state (Fig. \ref{fig:VO} d) shows that the additional charge is located on the vacancy site, forming an F-center. The formation of an F-center is not surprising as the reduction of $\mathrm{Mn^{2+}}$ to $\mathrm{Mn^{1+}}$ would be highly unfavourable. The markedly different strain dependence of the oxygen vacancy formation energy can thus be attributed to the formation of this F-center state. A more detailed study of different charge states and charge localisation will be presented elsewhere.

\section{Conclusions}

Using density functional calculations we have studied both Mn and O vacancies in epitaxially strained MnO thin films. We show that neutral $\mathrm{V_{Mn}^{''}}$ defects, which introduce two holes into the material, are significantly more favourable than charged defects in which these holes are filled with electrons. This is consistent with the fact that the additional electrons associated with charged defects occupy high-lying anti-bonding states. Strain has the expected effect on the formation energy of such a lattice contracting defect, making it more favourable under compression. Interestingly strain favours different arrangements of holes around the defect, which should lead to different anisotropic properties under slight compressive and tensile strains. Due to the formation of F-centers instead of charge localisation on the transition metals, oxygen vacancies are also favoured under compressive strain, in contrast to the simple chemical expansion model.

\section{Acknowledgements}

This work was financially supported by the ETH Z\"urich and by the ERC Advanced Grant program, No. 291151.

\bibliography{references.bib}

\end{document}